# Temperature Control using Fuzzy Logic

P. Singhala[1], D. N. Shah[2], B. Patel[3]

[1,2,3]Department of instrumentation and control, Sarvajanik College of Engineering and Technology Surat, Gujarat, INDIA

*Abstract*

*The aim of the temperature control is to heat the system up todelimitated temperature, afterwardhold it at that temperature in insured manner. Fuzzy Logic Controller (FLC) is best way in which this type of precision control can be accomplished by controller. During past twenty yearssignificant amount of research using fuzzy logichas done in this field of control of non-linear dynamical system. Here we have developed temperature control system using fuzzy logic. Control theory techniques are the root from which convention controllers are deducted. The desired response of the output can be guaranteed by the feedback controller.*

## 1. Introduction

Low cost temperature control using fuzzy logic system block diagram shown in the fig. in this system set point of the temperature is given by the operator using 4X4 keypad. LM35 temperature sensor sense the current temperature. Analog to digital converter convert analog value into digital value and give to the Fuzzy controller.

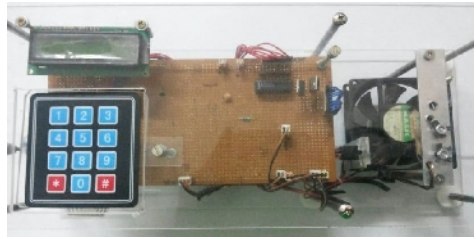

Fig: 1 Temperature control system

Controller calculates error between set point value and current value and consider as Input function of fuzzy logic. By fuzzification process controller calculate it membership. After in rule base and inference system output membership value calculated. Defuzzification process calculates actual value of PWM for heater and fan which is output of the temperature control system.

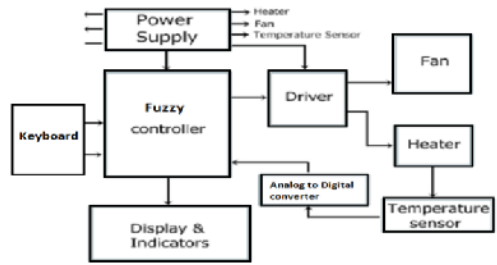

Fig 2: Block diagram of temperature control system

The process comprises of a heater, fan and a temperature sensor. The amount of current passing through the coil decides the temperature of the thin metal plate. Temperature detection of this metal plate can be done by dedicated temperature sensors. A fan is placed near to the heating mechanism. Amount of power delivered to both heater and fan can be controlled by passing a command through serial port via microcontroller. Now, microcontroller generate PWM (Pulse Width Modulation) signal for the MOSFET to deliver desired amount of power to fan and heater. It could thus be used as a small plant readily available for various experimentation and study purpose.

## 2. Working Principle

Temperature control system shown in below figure is works on the basic principle of fuzzy logic.The fundamentals of fuzzy logic elaborated by LotfiA.Zedeh, a professor at the University of California at Berkley.He presented fuzzy logic not as a control methodology, but as a method of processing data by allowing partial set membership instead of non membership. Until 70's due to insufficient small computer capability the method of settheory was not applied to control system.

Nonlinear mapping of an input data set to a scalar output data is known as fuzzy logic system. A fuzzy logic system consists of four main parts:

- Fuzzifier
- Rules
- inference engine
- defuzzifier.

These components and the general architecture of a fuzzy logic system are shown in Figure 3.

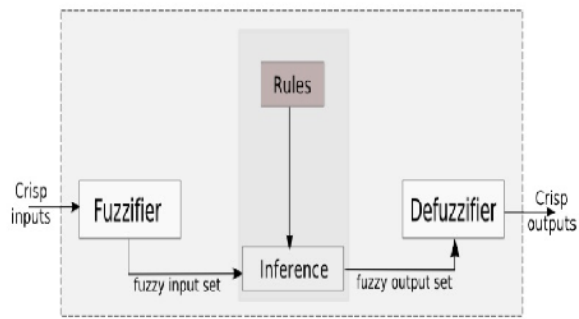

Fig 3: Fuzzy logic system

In order to exemplify the usage of a fuzzy logic system, consider a temperature control system controlled by a fuzzy logic controller. The temperature of the room can be adjusted by details like current temperature of the room and the target value by defined system. The comparison between the room temperature and the target temperature can be compared by fuzzy engine at certain period of time and produces a command of heating or cooling.

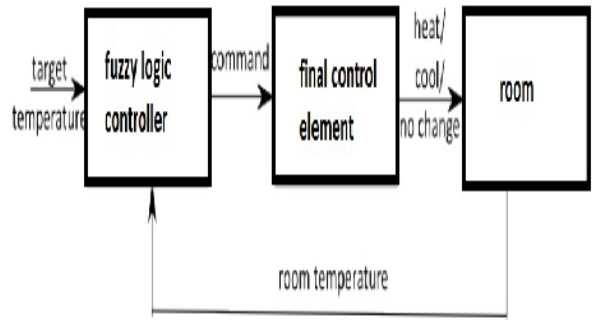

Fig 4: A simple fuzzy logic system to control room temperature

**Fuzzy logic algorithm:**

| Fuzzy logic algorithm | |
|---|---|
| 1 | Define linguistic variables and terms |
| 2 | Construct the membership function |
| 3 | Construct rule base |
| 4 | Convert crisp data to fuzzy values using the membership function |
| 5 | Evaluate rule in the rule base |
| 6 | Combine the result of each rule |
| 7 | Convert output data to non fuzzy values |

Table 1: Fuzzy logic algorithm

**Fuzzy set**

Before understand fuzzy set little terminology is necessary to understand.
**Set:** Objects having one or more similar characteristics can be collectedand classified into set.
**Member**: objects belonging to a set call member of the set.

In fuzzy set member have their membership grade associated with it. For example set of HOT temperature is decide between 60< TEMP< 80. If temperature is 60 degree then we say it is not belong to HOT set but in fuzzy logic it is belong to set but having membership grade 0. Similarly if temperature is 62 and 78 then it belong to fuzzy set with membership grade 0.10 and 0.90 respectively i.e. First member is not more likely belong to HOT set but 78 temperature is most likely belong to the set HOT.

**Membership Functions**

Implementation of membership function is vital in thefuzzification and defuzzification steps of a FLS, to evaluate the non-fuzzy input values to fuzzy linguistic terms and vice-versa. A membership function is implemented to measure the linguistic term. For instance, membership functions for the linguistic terms of temperature variable are plotted. The staggering feature of thefuzzy logic lies in fuzzification of the numerical value which need not to be fuzzified using only one membership function, so the value can be defined by various set at one particular instance. As below figure suggests, with different degree of memberships facilitate in such a way that a single temperature value can be considered in two different aspects at the single time.

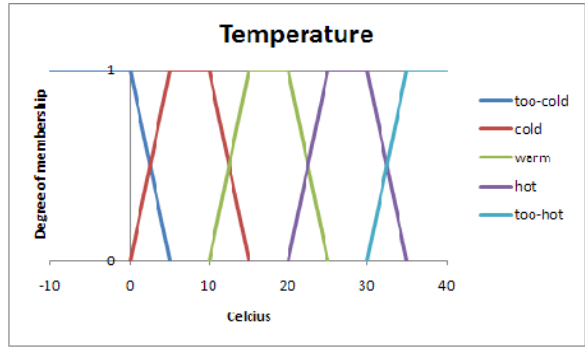

Fig 5: Membership Functions for T (temperature) = {too-cold, cold, warm, Hot, too-hot}

**Fuzzy Rules**

In a Fuzzy Logic, a rule base is constructed to control the output variable. A fuzzy rule is a simple IF-THEN rule with a condition and a conclusion. In Table 2, sample fuzzy rules for the temperature control system in Figure are listed.

| Fuzzy rules | |
|---|---|
| 1 | IF(temperature is cold OR too-cold)AND(target is warm)THEN command is heat |
| 2 | IF(temperature is hot OR too-hot)AND(target is warm)THEN command is cool |
| 3 | IF(temperature is warm)AND(target is warm)THEN command is heat |

Table 2: Sample fuzzy rules for temperature control system

Table 3 shows the matrix representation of the fuzzy rules for the said Fuzzy Logic.

| Temperature/target | Too-cold | Cold | Warm | Hot | Too-hot |
|---|---|---|---|---|---|
| Too-cold | No-change | Heat | Heat | Heat | Heat |
| Cold | Cool | No-change | Heat | Heat | Heat |
| Warm | Cool | Cool | No-change | Heat | Heat |
| Hot | Cool | Cool | Cool | No-change | Heat |
| Too-hot | Cool | Cool | Cool | Cool | No-change |

Table 3: matrix for the temperature control system

Row captions in the matrix contain the values that current room temperature can take, column captions contain the values for target temperature, and each cell is the resulting command when

International Journal of Instrumentation and Control Systems (IJICS) Vol.4, No.1, January 2014

the input variables take the values in that row and column. For instance, the cell (4, 5) in the matrix can be read as follows: If temperature is warm and target is hot then command is heat.
Fuzzification: Fuzzification is the process of making crisp quantity fuzzy. In real world, hardware such as digital voltmeter generates crisp data, but these data are subject to experimental error [8].

Defuzzification: The procedure of producing a quantitativeoutcome in fuzzy logic, given fuzzy sets and corresponding membership degrees can be described as term "Defuzzification". It is basicallyrequired in fuzzy control arrangements. These arrangements will contain large number of rules which willconvert a number of variables into a fuzzy result and eventually the converted variables called resultsareexpressed in terms of membership in fuzzy sets. To understand this, rules contrived to determine how much pressure to enforce might result in "Decrease Pressure (13%), Maintain Pressure (33%) and Increase Pressure (70%)". Defuzzification is rendering the membership degrees of the fuzzy sets into a particularconclusion in other words, real value [9].

## 3. Implementation of fuzzy logic

### Fuzzification of Input

In the fuzzification process, a real scalar value changes into a fuzzy value. Arrangements of Fuzzy variables ensurethat real values get translated into fuzzy values. After translating those real values into fuzzy values, the possible outcome is called "linguistic terms". The input linguistic variables for Fuzzy Logic Temperature Controller (FLTC) suggest two things. First it shows linguistically the difference between the set point and second it also express the measured and calculated signals from a temperature sensor. Input to FLTC is Error= (Set point-Temperature sensed). For fuzzified input, two functions including trapezoidal and triangular are used.To determine the range of fuzzy variables according to the crisp inputs is the primary requirement for proper running of the fuzzier program.Temperature difference which was sensed previously, is restricted to positive value. The following fuzzy sets are used: NEG =negative, SNEG=small negative, ZERO= zero, SPOZ=small positive, POZ= positive. Table suggests the Membership function for input linguistic variable. Membership function for input linguistic variable.

| No. | Crisp Input Range ( Error = Set Point – Current Temperature ) | Fuzzy Variable Name |
|---|---|---|
| 1 | -15 to -50 | NEG |
| 2 | 0 to +30 | SNEG |
| 3 | -15 to +15 | ZERO |
| 4 | 0 to +30 | SPOZ |
| 5 | +15 to +50 | POZ |

Table 4: Input linguistic variables

To include the linguistic variable negative (Neg) to a microcontroller, transformation of the pictorial representation into substantive code is needed.

### Fuzzy Membership Functions for Outputs

The output linguistic variables express linguistically the applied values to the FLTC actuators for temperature control. Present study considered typically one output variable, which is a PWM Wave for fan and Heater. In this case it is essential to attribute fuzzy memberships to yieldvariable, which has to be identical to the input variable. The fuzzy sets used for PWM Wave are as follows: Z = zero, L = large, M = medium, H = high, VH = very high.

| No | Fuzzy variable range output | Corresponding | Fuzzy variable name |
|---|---|---|---|
| 1 | 165.75 to 255 | 65% to 100% | VH |
| 2 | 127 to 204 | 50% to 80% | H |
| 3 | 165.75 to 89.25 | 65% To 35% | M |
| 4 | 127 to 51 | 50% to 20% | L |
| 5 | 89.25 to 0 | 35% to 0% | Z |

Table 5: Output linguistic variables

**Rule block**

Once the current values of the input variables are fuzzified, the fuzzy controllercontinues with the phase of "making decisions," or deciding what actions to take to bring the temperature to its set-point value. For the action to be initiated the measures are minimal time as well as minimal temperature. The restraintpolicy of a Fuzzy Control System is comprisedby the rule blocks.

In the rules 'IF' part depicts the situation, for which the rules are projected. The following 'THEN' part delineates the reaction of the fuzzy system in this state. The control policy of heater is structurally formulated according to fuzzy rules. For example, rule 1 "If error is NEG then firing angle is Z".

| No | Fuzzy variable range output | Corresponding | Fuzzy variable name |
|---|---|---|---|
| 1 | 165.75 to 255 | 65% to 100% | VH |
| 2 | 127 to 204 | 50% to 80% | H |
| 3 | 165.75 to 89.25 | 65% To 35% | M |
| 4 | 127 to 51 | 50% to 20% | L |
| 5 | 89.25 to 0 | 35% to 0% | Z |

Table 6: Fuzzy rules

**Defuzzification**

The outcome of defuzzification has to be in a numeric formso that it defines the PWM Wave of the MOSFET which is used toforce the fan and heater. Out of the number of ways to execute defuzzification; in thegiven scenario, weighted average defuzzificationis the best technique to obtain the crisp output. It can be further described by following equation (1).

$$D = \frac{\sum_{i=1}^{7} P(i) \times W(i)}{\sum_{i=1}^{7} W(i)} \ldots\ldots \qquad (1)$$

P[i] = Theextremum value of $i^{th}$ output membership function.
W[i] = The weight associated with $i^{th}$ rule.
The fuzzy variable can be converted into crisp output using C code fragment. One example of that is given below.

{

$$z = \frac{((f1 \cdot d1)+(f2 \cdot d2)+(f3 \cdot 127)+(f4 \cdot 89)+(f5 \cdot d5)+(f6 \cdot d6)+(f7 \cdot d7))}{(f1+f2+f3+f4+f5+f6+f7)};$$

PWM=z;
}

The degree of each membership function which was computed in the previous step of fuzzificationis encountered by the subprogram called "Defuzzify"and this after certain process it returns defuzzified output.This defuzzify output is employed to restraint the pulse width modulation wave of MOSFET.

## 4. Pulse width modulation technique

The temperature System has a Heater coil and a Fan. The heater assembly consists of an iron plate placed at a distance from a nichrome coil. When current passes throughthe coil it gets heated and in turn raises the temperature of the iron plate. We are interested to alter the heat generated by thecoil and also the speed at which the fan is operated. The amount of power which is to be delivered to fan and heater can be assured by several methods. We are using the PWM technique.

Modulation of the square wave which is in duty cycle is done by Pulse Width Modulation action.

Duty cycle = $T_{ON}$ / T
    Where;
$T_{ON}$ : ON time of the wave
T : Total time period of wave Power delivered to the load is proportional to T-ON time of the signal
This is used to control the current flowing through the heating element and also speed of the fan.

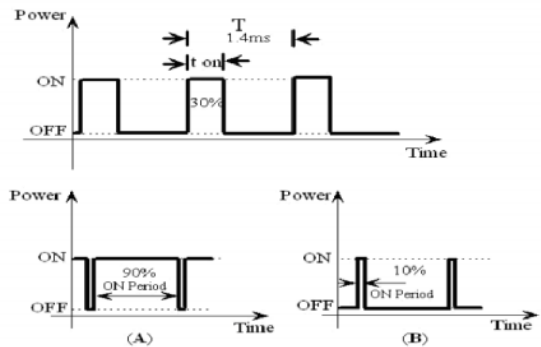

Fig 6: Pulse width modulation

An 80% duty cycle indicates that the fan is ON 80% of the time and OFF 20% of the time. The relation between the speed of the fan and the value of the applied pulse width modulation duty cycle is in direct proportion. In other words, a high duty cycle will produce high speeds and a low duty cycle will produce low speeds.

## 5. Outcome of temperature control using fuzzy logic

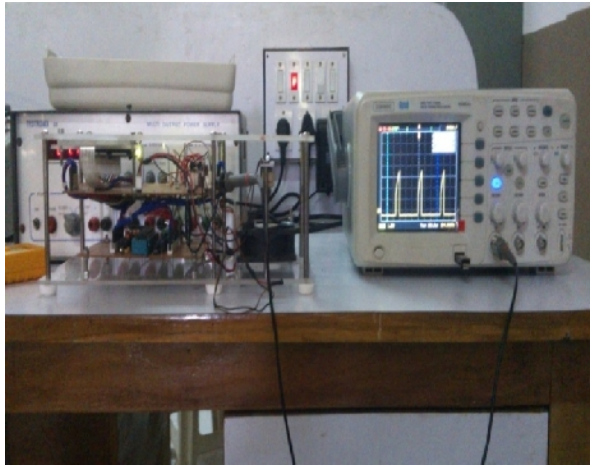

Fig 7: Overall view of system

The Temperature in the case study is as follow:

Set temperature: 45˙ C

Current temperature: 46˙ C

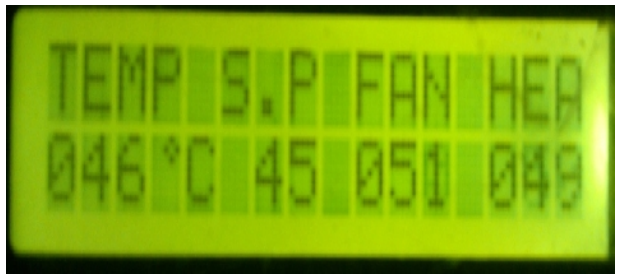

Fig 8: LCD display

Error = SP – CV = 45 - 46 = -1
Rule base Follow: Rule-2(SNEG) and Rule-3(ZERO)
Therefore Fuzzilization value f2=0.04
                        f3=0.933
Defuzzilization value $Z^{*} = 130.95$
The Duty Cycle of FAN speed = 51%
The Duty Cycle of HEATING COIL current = 49%

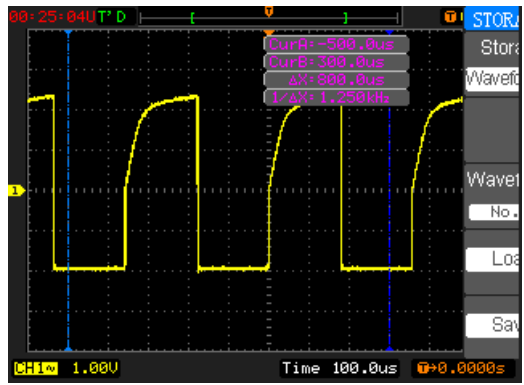

Fig 9: Graph of PWM

## 6. Conclusion

The fuzzy temperature controller is designed and implemented in microcontroller without using any special software tool. Unlike some fuzzy controllers with hundreds, or even thousands, of rules running on computer systems, a unique FLC using a small number of rules and simple implementation is demonstrated to solve a temperature control problem with unknown dynamics or variable time delays commonly found in industry. Also the final hardware is stand-alone system rather than a PC (personal computer/laptop computer) based system that takes control decision based on special software tools running on it and hence the design approach presented in this paper minimizes the total cost of hardware and software design. The control result can be improved by resizing the fuzzy sets and finer tuning for the membership functions.

Fuzzy Logic provides a completely different, unorthodox way to approach a control problem. This method focuses on what the system should do rather than trying to understand how it works. One can concentrate on solving the problem rather than trying to model the system mathematically, if that is even possible. This almost invariably leads to quicker, cheaper solutions. Once understood, this technology is not difficult to apply and the results are usually quite surprising and pleasing.

## Acknowledgement

We welcome this opportunity to express our heartfelt gratitude and regards to our professor Tejal Dave, Department of instrumentation & control Engineering, Sarvajanik college of engineering and Technology, Surat, for her unconditional guidance. She always bestowed parental care upon us and evinced keen interest in solving our problems. An erudite teacher, a magnificent person and a strict disciplinarian, we consider ourselves fortunate to have worked under her supervision. Without her co-operation, the extensive work involved in compiling background information and preparing the paper for publication would not be possible.would not be possible.

**Authors**

**PiyushSinghala**received B.E. degree in Instrumentation and Control Engineering from the Gujarat Technological University, India. He is right now working in a Precise Conchem Private Ltd. His area of interest covers Process Control as well as PLC and SCADA.

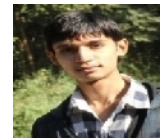

**Dhrumil Shah** received B.E. degree in Instrumentation and Control Engineering from the Gujarat Technological University, India. His research field covers basic concepts related to fuzzy logic and its application.

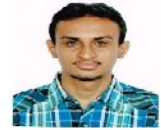

**Bhavikkumar Patel** has received B.E. degree in Instrumentation and Control Engineering from Gujarat Technological University, India. He also seeks some of his interests in control system designing and automation.

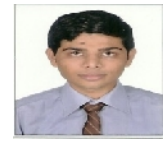